\documentclass[preprint,12pt]{elsarticle}
\usepackage[english]{babel}
\usepackage{amsmath}
\journal{Physics Letters B}

\begin{document}
\begin{frontmatter}
\title{Brane bounce-type configurations in a string-like scenario.}

\author[ufc,ifce]{L. J. S. Sousa}
\ead{luisjose@fisica.ufc.br}

\author[ifpb]{C. A. S. Silva}
\ead{calex@fisica.ufc.br}

\author[ufc]{C. A. S. Almeida}
\ead{carlos@fisica.ufc.br}

\address[ufc]{Departamento de F\'{\i}sica - Universidade Federal do Cear\'{a} - UFC \\  C.P. 6030, 60455-760 Fortaleza - Cear\'{a} - Brazil
}

\address[ifce]{Instituto Federal de Educa\c{c}\~{a}o Ci\^{e}ncia e Tecnologia do Cear\'{a} (IFCE) - Campus de Canind\'{e} \\
62700-000 Canind\'{e} - Cear\'{a} - Brazil}

\address[ifpb]{Instituto Federal de Educa\c{c}\~{a}o Ci\^{e}ncia e Tecnologia da Para\'{i}ba (IFPB) - Campus Campina  \\
Rua Tranquilino Coelho Lemos, 671, Jardim Dinam\'{e}rica,\\ Campina Grande - Para\'{\i}ba - Brazil}

\begin{abstract}

Brane world six dimensional scenarios with string like metric has been proposed to alleviate the problem of field localization.
However, these models have been suffering from some drawbacks related with energy conditions
as well as from difficulties to find analytical solutions. In this work, we propose a model where a brane is made of a scalar field with bounce-type configurations and embedded in a bulk with a
string-like metric. This model produces a sound AdS scenario where none of the important physical quantities is infinite. Among these quantities
are the components of the energy momentum
tensor, which have its positivity ensured by a suitable choice of the bounce configurations.
Another advantage of this model is that the warp factor can be obtained  analytically from the equations of motion for the
scalar field, obtaining as a result a thick brane configuration, in a six dimensional context. Moreover, the study of the scalar field localization
in these scenario is done.

\end{abstract}


\end{frontmatter}

\section{Introduction}

The suggestion that our world is a three brane embedded in a higher-dimensional space-time has attracted the attention of the physics community in the
last years. It is basically because the brane world idea has brought solution for some intriguing problems in the Standard Model, like the hierarchy
problem \cite{I.Antoniadis1998, N.Arkani-Hamed1999, N.Arkani-Hamed1999a, Randall1999a,Randall1999}. Another important possibility, that seems to be open in brane models, is that of explaining the smallness of the observed cosmological
constant. The fact that extra space-time dimensions can introduce extra contributions to the vacuum energy that can allow for a vanishing four
dimensional cosmological constant was observed some years ago independently of branes \cite{va.rubakov-plb125, gr.dvali-nob504}. Branes on the other
hand allow for an interplay between higher dimensional and four dimensional cosmological constant contributions. Such a self-tuning mechanism has been
pointed out some years ago \cite{ s.kachru-prd62}. However, the solutions found contain many naked singularities.

The mainly kinds of theories that carrier the brane world basic idea are the one first proposed by Arkani-Hamed, Dimopoulos and Dvali
\cite{I.Antoniadis1998, N.Arkani-Hamed1999, N.Arkani-Hamed1999a} and the so called Randall - Sundrum model \cite{Randall1999a,Randall1999}.
In the last, it is assumed that, in principle, all the matter fields are constrained to propagate only on the brane, whereas gravity is free to propagate
in the extra dimensions \cite{p.kanti-ijmpa}. 

However the assumption that the Standard Model particles are initially trapped on the brane is not so obvious in this framework.
In this way, among the main issues
approached in the brane world context, is the problem of localization of several fields and resonances in such branes. The importance of this
subject comes from the fact that, if indeed present, the extra dimensions will inevitably change our notion for the
universe. The introduction of extra dimensions affects both gravitational
interactions and particle physics phenomenology, and leads to modifications in standard
cosmology.  In this way, the investigation of field localization issue can guide us to which kind
of brane structure is more acceptable phenomenologically \cite{W.T.Cruz2009}, which makes
interesting to look for an alternative field theoretic localization mechanism in brane world scenarios \cite{Oda2000}.

Several ideas and generalizations have been proposed in order to approach this issue. Among these ideas, a smooth generalization of the Randall-Sundrum scenario has been proposed in \cite{Kehagias2001}, where five-dimensional gravity is coupled to scalar
fields. This generalization gives rise to a new class of brane models, which are now known as ``thick branes''(A detailed review in this
subject may be found in Dzhunushaliev \cite{Dzhunushaliev2010}. According to this author the first work in the subject that we call ``thick brane''
today were done by \cite{Akama1983, Rubakov1983}). It has been shown that thick branes consists in a more realistic model that the Randal-Sundrum
one, since no singularities appear in this approach due to the form of the scalar potential functions.

Along the years, new models have been proposed for such
branes with internal structures, constructed with one \cite{D.Bazeia1} or more \cite{D.Bazeia2, m.eto, a.desouza} scalar fields in a five dimensional
scenario.
In these works, thick brane models,
gravitons and fermions, as well as gauge fields can be localized on the brane. However, gauge fields are localized only with the help of the dilaton
field. The Kalb-Ramond field localization in this scenario was also studied by \cite{Cruz2009}. There the use of the dilaton was again necessary in
order to localize of the Kalb-Ramond field on the brane.
On the other hand, other scenarios have been proposed where thick brane solutions are extended to spacetimes with dimension more than
five \cite{Dzhunushaliev2010}. Among these works, we have some where branes are embedded in a bulk with a string-like metric.
The mainly motivation to
study branes in the presence of a string-like bulk comes from the fact
that most of the Standard Model fields are localized on a string-like defect. For example, spin-$0$, spin-$1$, spin-$2$, spin-$1/2$ and spin-$3/2$ fields are
all localized on a string-like. Particularly, the bosonic fields are localized with exponentially decreasing warp factor, and the fermionic fields are
localized on defect with increasing warp factor \cite{Oda2000}. Even more interesting is the fact that spin-$1$ vector \cite{Oda2000},
as well as the Kalb-Ramond field \cite{W.T.Cruz2009}, which are not localized on a domain wall, in Randal-Sundrum model, can be localized in the
string-like defect.

However, most of the thick brane models in six dimensional scenarios, proposed so far, have been suffering from some drawbacks. The first difficult is
related with the introduction of scalar fields as a matter-energy source in the equations. In this case it is very difficult to find analytical
solution to the scalar field, and the warp-factor. Koley and Kar \cite{Koley2007} have suggested a model where analytical solutions can be found in a six
dimensional scenario, however they run into a second difficult. This difficult is related with the positivity of the components of the
energy-momentum tensor and has been found by other authors too \cite{Dzhunushaliev2010, Dzhunushaliev2008, Koley2007}. Finally, problems with field localization was found at least in
one case \cite{Dzhunushaliev2008}.

In this work, we propose a model where a brane is made of a scalar field with bounce-type configurations and embedded in a bulk with a
string-like metric. This model produces a sound AdS scenario where none of the important physical quantities is infinite. Among these quantities
are the components of the energy momentum
tensor, which have its positivity ensured by a suitable choice of the bounce configurations.
Another advantage of this model is that the warp factor can be obtained  analytically from the equations of motion for the
scalar field, obtaining as a result a thick brane configuration, in a six dimensional context. Moreover, the study of the scalar field localization
in these scenario is done.

This paper is organized as follows. In the section 2, we introduce a model where a bulk scalar
field with bounce-type configurations generates a brane which is embedded in a bulk with a string-like metric.
In section 3, we investigate the possibility of field localization in the scenario introduced in the section 2.
Section 4 is devoted to conclusions.

\section{The model} \label{section2}

In this section, we will introduce the basic ideas of this work. In this way, we will construct a model where a bulk scalar field with bounce-type configurations generates a brane which is embedded in a bulk with a
string-like metric. The use of bulk scalar fields to generate branes was introduced by
\cite{wd.goldberger-prl83, wd.goldberger-prd60}, and has been largely studied in the literature
\cite{o.dewolfe-prd62, rn.mohapatra-prd62, p.kanti-plb481, jm.cline-prd64, jm.cline-plb495, a.flachi-npb610}. In the six-dimensional context,
we highlight the work done by Koley and Kar \cite{Koley2007}, where the brane is made of scalar fields and the authors found analytical
``thin brane'' solutions. In this work, the authors dealt with two different models.
In the first one, the presence of a bulk phantom scalar field was supposed. In the second one,
it was supposed the presence of a bulk Brans-Dicke scalar field. Several progress have been obtained in that work in the intend of construct
brane solutions in six dimensions, as well as, in the task of localize physical fields. Among these results, is the localization of massless spin
fields ranging from $0$ to $2$ on a single brane by means of gravity only. Moreover, in this model, the sixth dimension seems to facilitate the
localization of vectors fields, a result which does not exist in five dimensions. However, some troubles with the energy conditions
(WEC, SEC, NEC) \cite{m.visser-aip} were found. In the scenario introduced by Koley and Kar, the energy momentum tensor violates all the energy
conditions since its components are not positive defined. Among the bad consequences
of this, we have that the bulk spacetime obtained in that setup could be not dynamical stable. The authors tried to release the violation of
the energy conditions saying that it also occurs in the Randall-Sundrum model \cite{Randall1999a}, however the problem remains open.

In our work, we will try to overcome the problems with the energy conditions by using a scalar field model with bounce-type configurations
in a string-like scenario. In the same way of the model introduced by Koley and Kar, the model we will introduce here has the advantages to be
analytical. However,
the introduction of the bounce-type configurations to the scalar field that generates the brane will solve the problems with the energy conditions.
Moreover, as we will show, a sound scenario for field localization is produced.

To begin with, we will assume a six dimensional action for a bulk scalar field in a potential $V(\phi)$ minimally coupled to gravity in the presence
of a cosmological constant:
\begin{equation}
S=\frac{1}{2\kappa_{6}^{2}} \int d^{6}x \sqrt{-\;^{(6)}g}\Big[(R - 2\Lambda) + g^{AB}\nabla_{A} \phi \nabla_{B}\phi - V(\phi)\Big]\;, \label{action}
\end{equation}
where $\kappa _{6}$ is the $6$-dimensional gravitational constant, and $\Lambda$ is the bulk cosmological constant.

The equations of motion obtained by variation of the action \eqref{action} are
\begin{eqnarray}
R_{MN}-\frac{1}{2} g_{MN}R  &=&  \kappa_{6}^{2} \Big[\partial_{M} \phi \partial_{N} \phi - g_{MN}\Big(\frac{1}{2}(\partial \phi)^{2} + V(\phi) \Big)\Big] \nonumber\\
& - & \Lambda g_{MN}
\end{eqnarray}
and
\begin{equation}
\frac{1}{\sqrt{-\;^{(6)}g}}\partial_{M} \Big\{ \sqrt{-\;^{(6)}g}\;g^{MN} \partial_{N}\phi \Big\} = \frac{\partial V}{\partial \phi}\;.
\end{equation}

We have that, in the absence of gravity, for a scalar potential of the double well type $V(\phi) = \frac{\lambda}{4}(\phi^{2} -v^{2})^{2}$, the scalar field equation possess bounce-like statics solutions depending only on the radial extra dimension, where the simplest of which is
\begin{equation}
\phi(r) = v \tanh(ar) \label{scalar-field}\; ,
\end{equation}
with $a^{2} \equiv \lambda v^{2}/2$.

Now let us introduce the string-like metric
\begin{eqnarray}
\lefteqn {ds^{2}=g_{MN}dx^{M}dx^{N}}\nonumber\\
& &=g_{\mu\nu}dx^{\mu}dx^{\nu}+\tilde{g}_{ab}dx^{a}dx^{b}\nonumber\\
& &=e^{-A(r)}\hat{g}_{\mu\nu}dx^{\mu}dx^{\nu}+dr^{2}+e^{-B(r)}d\Omega_{(5)}^{2}\;, \label{sl-metric}
\end{eqnarray}
where $M,N,...$ denote the $6$-dimensional space-time indices, $\mu, \nu, ...,$ the $4$-dimensional brane ones, and $a, b, ...$ denote the $2$-extra spatial dimension ones.

From the equations above, we obtain the following field equations for the Einstein-scalar system:
\begin{equation}
e^{A(r)} \hat{R} - 2 A'(r) B'(r) - 3 (A'(r))^{2} = - 2 \kappa_{D} ^{2} t_{r} + 2\Lambda \label{eq1}\; ,
\end{equation}
\begin{eqnarray}
e^{A(r)} \hat{R} - 5A'(r)^{2} + 4A''(r) = - 2 \kappa_{D} ^{2} t_{\theta} + 2 \Lambda \label{eq2}\; ,
\end{eqnarray}
and
\begin{eqnarray}
\frac{1}{2} e^{A(r)} \hat{R} &+& 3 A''(r)   - \frac{3}{2} A'(r) B'(r) \nonumber \\
&-& 3 (A'(r))^{2} + B''(r)
- \frac{1}{2}B'(r)^{2} = - 2 \kappa_{D} ^{2} t_{0} + 2 \Lambda \; ,
\end{eqnarray}
where $t_{i} (i = 0, r, \theta)$ are functions of $r$, and are given by the non-vanish components of the energy-momentum tensor $T_{MN}$
($T^{\mu}_{\nu} = \delta^{\mu}_{\nu}t_{0}(r)$, $T_{r} ^{r} = t_{r}(r)$, $T_{\theta} ^{\theta} =t_{\theta}(r)$):
\begin{equation}
t_{0}(r) = t_{\theta}(r) =  - \frac{\phi'^{2}}{2} + V(\phi)  \; ,
\end{equation}
\begin{equation}
t_{r}(r) = \frac{\phi'^{2}}{2} + V(\phi).
\end{equation}
Note that with this form to the energy-momentum tensor, we keep spherical symmetry.

In addition, the scalar curvature is given by
\begin{equation}
R = -5(A'(r))^{2} - 2A'(r)B'(r) - \frac{1}{2}(B'(r))^{2} + 4A''(r) + B''(r)\;.
\end{equation}

\begin{figure}[htb]
 \centering 
 \fbox{\includegraphics[width=8.0cm,height=4.5cm]{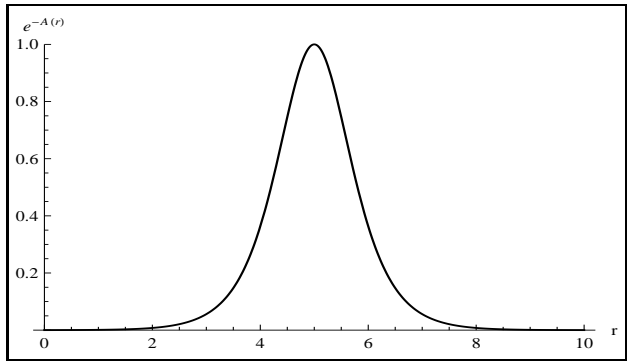}}
 \caption[Figure 1:]{$e^{-A(r)}$ profile for $\beta = 1; a = 1$}
 \label{fig:1}
 \end{figure}

From now on, we will restrict us to the case where $B(r) = A(r)$. Then, integrating twice the sum of Eq.\eqref{eq1} and Eq.\eqref{eq2}, we obtain for a scalar field given by Eq.\eqref{scalar-field},
the metric exponent function (this solution ensures $A(0)= A^{'}(0) = 0)$
\begin{equation}
A(r) = \beta \ln \cosh(ar) + \frac{\beta}{2} \tanh^{2}(ar) \label{warp-factor}
\end{equation}
with $\beta = \frac{1}{3} \kappa_{D} ^{2} \nu ^{2}$. A profile of the warp factor $e^{-A(r)}$ is given in Fig. \eqref{fig:1}.
This profile ensures the finiteness of the relation between the four ($M_{p}$) and six ($M_{6}$) dimensional reduced Plank scale \cite{Gherghetta2000}
\begin{equation} \label{Plank-scale}
M_{p} ^{2} = 2 \pi M_{6} ^{4} \int_{0} ^{\infty} {dr e^{-3/2 A(r)}} \; .
\end{equation}

\begin{figure}[htb] 
       \begin{minipage}[b]{0.48 \linewidth}
           \fbox{\includegraphics[width=\linewidth]{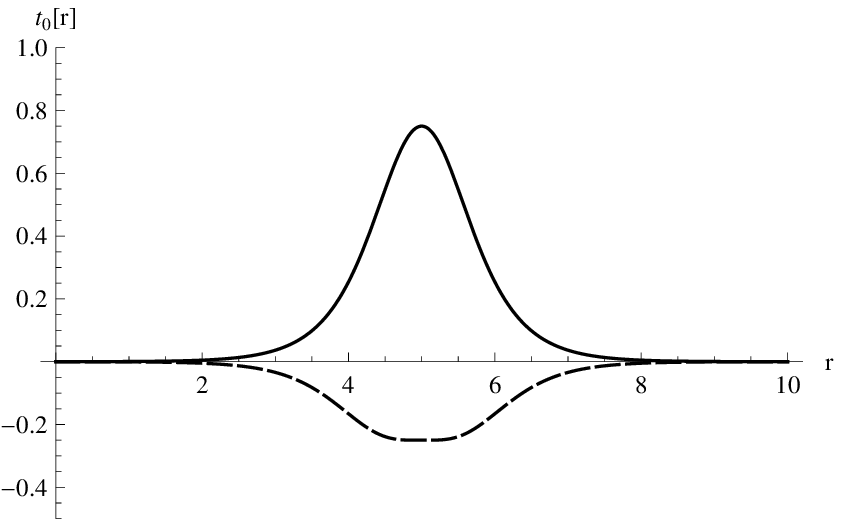}}\\
           \caption{\it $t_{0}(r) = t_{\theta}(r)$ profile for $v = -1$ (filled line), and for $v = 1$ (dashed line) }
           \label{fig:2}
       \end{minipage}\hfill
       \begin{minipage}[b]{0.48 \linewidth}
           \fbox{\includegraphics[width=\linewidth]{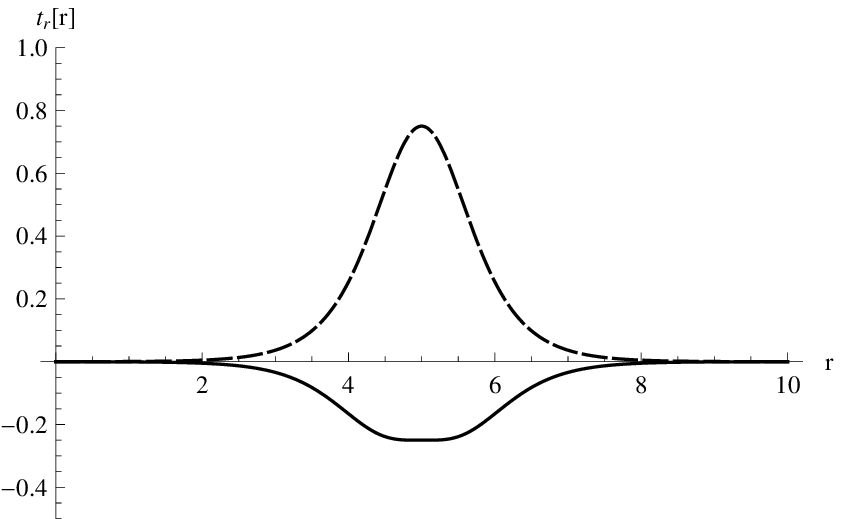}}\\
           \caption{\it $t_{r}(r)$ profile for $v = 1$ (filled line), and for $v = - 1$ (dashed line)}
           \label{fig:3}
       \end{minipage}
   \end{figure}

\begin{figure}[htb]
 \centering 
 \fbox{\includegraphics[width=8.0cm,height=4.5cm]{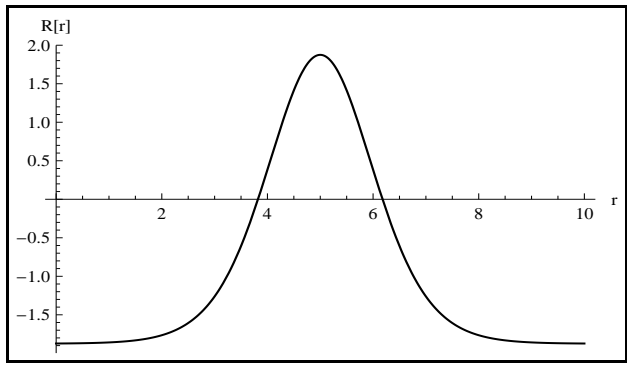}}
 \caption[Figure 1:]{R profile for $\beta = 1; a = 0.5$}
 \label{fig:4}
 \end{figure}

In order to have a physically accepted scenario, it is necessary that the energy momentum components and the curvature scalar be finite.
To analyze the behavior of these quantities, we have plotted the components $t_{0}(r) = t_{\theta}(r)$, $t_{r}(r)$ of the energy momentum tensor,
and the scalar curvature R in Fig. \eqref{fig:2},
Fig. \eqref{fig:3}, and Fig. \eqref{fig:4}, respectively. The energy-momentum components depends only on the scalar field derivative and the scalar field potential.
The figures show that the model proposed in this work produces a sound scenario where none
of this important
quantities is infinite, and the positivity of energy-momentum tensor components is ensured with a suitable choice of the
scalar field constants. In other words, it is the bounce-type configuration
that ensures that the model is physically acceptable. Besides, as we can see, the curvature scalar profile reveals an AdS scenario, since R is asymptotically negative.

In this way, differently from the results found by Koley and Kar, our energy density may be positive or negative on the brane depending on the choice of the bounce configurations, in
a way that, the problems with energy conditions can be circumvented. Moreover, the warp factor that we found
is equal to 1 at $r = 0$ which ensures that on the brane one has a 4D Minkowski space-time. Moreover, as $r$ goes to
zero or infinity, our warp factor goes to 0, as can be seen in Figure \eqref{fig:1}.

It is also interesting to point out the possibility to localize all the standard model fields in this model. We know that in five
dimensions it is possible to localize chiral fermions in the ``5D version of this model'' \cite{Kehagias2001}.
However, to localize vector field in this set up, in five dimensions, we need to have a dilaton field present in the model forming a
``bounce-gravity-dilaton system" \cite{Kehagias2001}. This procedure is using to localize the Kalb-Ramond field that is not localized
only by means of gravity in this background \cite{Tahim2009}. In six dimensions the situation changes and it is possible to localize
either the vector field \cite{Oda2000, Oda2000a} and the Kalb-Ramond field \cite{Luis-ip} without the necessity of the dilaton field,
in AdS Randall-Sundrum model. In this same context the fermionic fields are localized \cite{Liu2007}.
We expect, in this way, to localize
fields in this scenario that is more realistic than the RS model ones, without the necessity of the dilaton field.
In this work, we will give the first step in this analysis
considering the scalar field.

\section{Scalar field localization} \label{section3}

The investigation of field localization issue can guide us to which kind
of brane structure is more acceptable phenomenologically \cite{W.T.Cruz2009}.
The first natural step in the investigation on the possibility to localize fields in any braneworld scenario is try to localize the zero mode
of a scalar field. In this way, in order to study the localization of the scalar field in this context,
we will begin with the equations of motion for the scalar field in six dimensions
\begin{equation}
\frac{1}{\sqrt{-g}}\partial_{M}\left( \sqrt{-g}g^{MN}\partial_{N}\Phi\right) = 0 \;.
\end{equation}
By separating the brane coordinates from the extra coordinates ones, we simplify the equation above and get
\begin{equation} \label{eq14}
e^{A(r) - B(r)/2} \eta^{\mu\nu}\partial_{\mu}\partial_{\nu}\Phi + \partial_{r}\left( e^{-2A(r)
- B(r)/2}\partial_{r}\Phi\right) + \frac{e^{2A(r) - B(r)/2}}{R_{0} ^{2}} \partial_{\theta} ^{2} \Phi = 0\;,
\end{equation}
where $\eta^{\mu\nu}$ is the metric of the quadri-dimensional Minkowski space-time.

If we assume the following decomposition for the scalar field
\begin{equation}
\Phi (x^{M})= \phi(x^{\mu})\sum_{lm}{\chi_{m}(r)e^{il\theta}}\;,
\end{equation}
we can separate the variables in the equation \eqref{eq14}. Then, by requiring that $\eta^{\mu\nu}\partial_{\mu} \partial_{\nu}\phi = m^{2}\phi$, we get the following equation
for the radial variable
\begin{equation}
e^{A(r) + B(r)/2}\partial_{r}\left[ e^{-2A(r) - B(r)/2}\partial_{r}\chi(r)\right] + \left[m^{2} - \frac{l^{2}e^{B(r) - A(r)}}{R_{0} ^{2}} \right]\chi(r)  = 0
\end{equation}
or, yet
\begin{equation} \label{eq17}
\chi''(r) - \left( 2A'(r) + \frac{B'(r)}{2}\right)\chi '(r) + \left[ m^{2}e^{A(r)} - \frac{l^{2} e^{B(r)}}{R_{0} ^{2}} \right] \chi(r)  = 0\;,
\end{equation}
where the prime means the derivative with respect to $r$.

To solve this equation, we proceed by changing both the dependent and independent variables in order to obtain
a Schr\"{o}dinger like equation. So if we assume $z'(r) = e^{A(r)/2}$, we get
\begin{equation} \label{eq18}
\ddot{\chi}(z) - \left( \frac{3 \dot{A}(z)}{2} + \frac{\dot{B}(z)}{2}\right)\dot{\chi} (z) + \left[ m^{2}
- \frac{l^{2} e^{B(z) - A(z)}}{R_{0} ^{2}} \right] \chi(z)  = 0\;,
\end{equation}
where the point means the derivative with respect to $z$.

If we take $\chi(z) = \Omega(z)\Psi(z)$ with $\Omega(z) = \Omega_{0} e^{(3A(z) + B(z))/4}$, where $\Omega_{0}$ is an integration constant, we will have
\begin{equation}
 -\frac{d^{2}\Psi(z)}{dz^{2}} + V(z)\Psi(z) = m^{2}\Psi(z)\;,
\end{equation}
where
\begin{equation} \label{eq20}
V(z) = \left[ \frac{3\dot{A}(z) + \dot{B}(z)}{4}\right]^{2} - \left[ \frac{3\ddot{A}(z) + \ddot{B}(z)}{4}\right]
+ \frac{l^{2}}{R_{0} ^{2}}e^{B(z) - A(z)} \;.
\end{equation}

In the case where $A\equiv B$, expression \eqref{eq18} is simplified to
\begin{equation}
\ddot{\chi}(z) - 2\dot{A}(z) \dot{\chi} (z) + \left[ m^{2} - \frac{l^{2}}{R_{0} ^{2}} \right] \chi(z)  = 0\;.
\end{equation}

Moreover, the respective Schr\"{o}dinger-like equation and potential are given by
\begin{equation} \label{eq22}
 -\frac{d^{2}\Psi(z)}{dz^{2}} + V(z)\Psi(z) = m^{2}\Psi(z)
\end{equation}
with
\begin{equation} \label{eq23}
V(z) = \dot{A}(z)^{2} - \ddot{A}(z) + \frac{l^{2}}{R_{0} ^{2}}\;.
\end{equation}

\begin{figure}[htb]
 \centering 
 \fbox{\includegraphics[width=8.0cm,height=4.5cm]{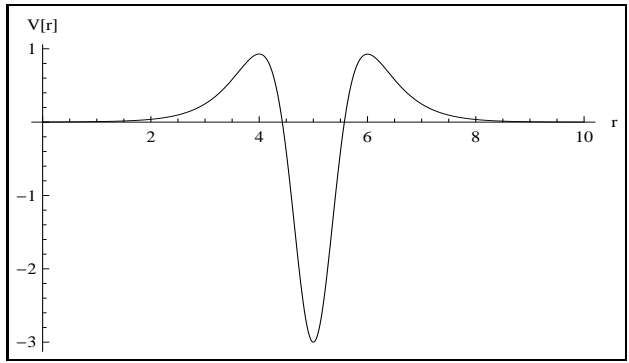}}
 \caption[Figure 1:]{V(r) profile for $\beta = 2; a = 1$}
 \label{fig:5}
 \end{figure}

In terms of the $r$ derivatives, the potential \eqref{eq23} reads
\begin{equation}
V(r) = e^{-A(r)}\left[ \frac{3A'(r)^{2}}{2} - A''(r) \right] + \frac{l^{2}}{R_{0} ^{2}} \; .
\end{equation}
This is a volcano potential, as can be seen in Fig.\eqref{fig:5}.
This kind of potential is very common in the literature, in the context of brane models and field localization,
and it is important to ensure localization.

Now, we will turn back to equation \eqref{eq17} to study the so called zero mode $m = 0$ and s-wave $l = 0$. In this
 case equation \eqref{eq17} is reduced to
\begin{equation}
\chi''(r) - \left( 2A'(r) + \frac{B'(r)}{2}\right)\chi '(r)  = 0\;.
\end{equation}
This equation admits, as unique finite solution, the trivial solution $\chi_{0} = constant$. So we will place this solution in the scalar action
\begin{equation}
S = -\frac{1}{2}\int{d^{6}x \sqrt{-g}g^{MN}\partial_{M}\Phi\partial_{N}\Phi} \; .
\end{equation}
For the case at hands, this action is given by
\begin{equation}
S_{0} = -\frac{1}{2}\int{d^{6}x R_{0} e^{-A(r) - B(r)/2}\eta^{\mu\nu}\partial_{\mu}\Phi_{0}\partial_{\nu}\Phi_{0}}\; .
\end{equation}
The interesting integral here is
\begin{equation}
I \propto \int_{0} ^{\infty}{dr e^{-A(r) - B(r)/2}} \; .
\end{equation}

The possibility of field localization is insured by the finiteness of this integral. In this way,
it is sufficient that $A(r) + B(r)/2 > 0$ ( in the case where $A(r) = B(r)$, we only need $A(r) > 0$) in order to have zero mode localization for the scalar field.
One easily can see that the form \eqref{warp-factor} to $A(r)$ obeys this condition.
This result shows that we have zero mode localization for the scalar field. It is interesting to
note that any non-gravitational trapping mechanism has not been necessary to localize scalar field
in this model, which is an advantage when compared with results of Dzhunushaliev and Folomeev \cite{Dzhunushaliev2008}.

\section{Remarks and conclusions} \label{section4}

We have constructed a model where a thick brane is generated from a scalar field on a string-like defect.
The model has similar characteristics to the one encountered by Dzhunushaliev and Folomeev \cite{Dzhunushaliev2008}, but in
our case we have the advantage that our model has analytical solutions. The warp factor that we found
is equal to 1 at $r = 0$. It ensures that we have 4D Minkowski space-time on the brane. As $r$ goes to
zero or infinity our warp factor goes to 0, as can be seen in Figure \eqref{fig:1}.

Differently from the results found in ref. \cite {Dzhunushaliev2008} our energy density may be positive or negative on the brane and is asymptotically
zero when $r$ goes to zero or infinity, as the warp factor. The negativity of the energy density may be used to
explain the formation of the brane where the repulsion from the negative energy density can balance the attraction
from gravity. Additionally the energy density was derived from a scalar field and it was possible to find an analytical
solution to the warp factor.

Another work where scalar fields were used
to construct brane in six dimensions was done by Koley and Kar \cite{Koley2007}. The authors found analytical
``thin brane'' solutions to the warp factor from scalar fields. However some problems with the energy conditions
were found. Our solution, in contrast, is an AdS type solution which presents a energy density that may be negative,
zero or positive depending on the choice of the bounce configurations. It prevents our model from problems with the energy
conditions (WEC, SEC, NEC) \cite{m.visser-aip} that is encountered in Ref. \cite{Koley2007} .

Moreover, in the context of braneworld it is suitable to investigate if a model is able to localize fields, in general.
In order to analyze if our solution enables field localization, we studied the zero mode scalar field localization.
Our results show that we have zero mode localization for the scalar field. It is interesting to note that
any non-gravitational
trapping mechanism has not been necessary to localize scalar field in this model which is an advantage when compared with \cite{Dzhunushaliev2008}.
In future works we will study the localization of other fields in this context.

The authors would like to thank CNPq and CAPES (Brazilian agencies) for financial support.


\begin{thebibliography}{99}

\bibitem{I.Antoniadis1998} I. Antoniadis, N. Arkani-Hamed, S. Dimopoulos, and G. Dvali, New dimensions at a millimeter to a Fermi and superstrings
at a TeV, Phys.Lett. B 436 (1998) 257.

\bibitem{N.Arkani-Hamed1999} N. Arkani-Hamed, S. Dimopoulos, and G. Dvali, The Hierarchy problem and new dimensions at a millimeter,
Phys. Lett. B 429 (1998) 263.

\bibitem{N.Arkani-Hamed1999a} N. Arkani-Hamed, S. Dimopoulos, and G. Dvali, Phenomenology, astrophysics and
cosmology of theories with submillimeter dimensions and TeV scale quantum gravity, Phys. Rev. D 59 (1999) 086004.

\bibitem{Randall1999a} L. Randall and R. Sundrum, An alternative to compactifcation. Physical Review Letters 83 (1999) 4690.

\bibitem{Randall1999} L. Randall and R. Sundrum, Large mass hierarchy from a small extra dimension. Physical Review Letters 83 (1999) 3370.

\bibitem{va.rubakov-plb125} V.A.~Rubakov and M.E.~Shaposhnikov, Do We Live Inside a Domain Wall? Phys. Lett.  B 125 (1983) 136.

\bibitem{gr.dvali-nob504} G.R.~Dvali and M.A.~Shifman, Dynamical compactification as a mechanism
of spontaneous supersymmetry breaking,  Nucl. Phys.  B 504 (1997) 127.

\bibitem{s.kachru-prd62} S.~Kachru, M.B.~Schulz and E.~Silverstein, Selftuning flat domain walls in
5-D gravity and string theory, Phys. Rev. D 62 (2000) 045021.

\bibitem{p.kanti-ijmpa} P. Kanti. Black holes in theories with large extra dimensions: a review.
International Journal of Modern Physics A 19 (2004) 4899.

\bibitem{W.T.Cruz2009} W. T. Cruz, A. R. Gomes, C. A. S. Almeida, Resonances in gravitational scenario given
by deformed branes, Eur. Phys. J. C71 (2011) 1709.

\bibitem{Oda2000} I. Oda. Localization of matters on a string-like defect. Physics Letters B 496 (2000) 113.

\bibitem{Kehagias2001} A Kehagias and K. Tamvakis, Localized gravitons, gauge bosons and
chiral fermions in smooth spaces generated by a bounce, Phys. Lett. B, 504 (2001) 38.

\bibitem{Dzhunushaliev2010} V. Dzhunushaliev, V. Folomeev,and M. Minamitsuji, Thick brane solutions. Rep. Prog.
Phys. 73  066901 (2010) [arXiv:0904.1775 [gr-qc]].

\bibitem{Akama1983} K. Akama, Pregeometry, in: K. Kikkawa, N. Nakanishi and H. Nariai, Lecture Notes in Physics, Springer-Verlag, 1983,
pp. 267-271; (A TeX-typeset version is also available in e-print hep-th/0001113).

\bibitem{Rubakov1983} V.A. Rubakov and M.E. Shaposhnikov, Phys. Lett. B125 (1983) 136.

\bibitem{D.Bazeia1} D. Bazeia, C. Furtado, A.R. Gomes, Brane structure from scalar field in warped space-time, JCAP 02, 002 (2004) 0405:012.

\bibitem{D.Bazeia2} D. Bazeia , A.R. Gomes, Bloch brane, JHEP 0405 (2004) 012.

\bibitem{m.eto} M. Eto and N. Sakai, Solvable models of domain walls in N = 1 supergravity, Phys. Rev. D 68 (2003) 125001

\bibitem{a.desouza} A. de Souza Dutra, A. C. Amaro de Faria Jr., M. Hott, Degenerate and critical Bloch branes, Phys. Rev. D 78 (2008) 043526.

\bibitem{Cruz2009} W. T. Cruz, M. O. Tahim, C. A. S. Almeida, Results in Kalb-Ramond field localization and resonances on deformed branes,
Europhysics Letters 88 (2009) 41001.

\bibitem{Koley2007} Ratna Koley and Sayan Kar, Braneworlds in six dimensions: new models with bulk scalars,
Class. Quantum Grav. 24 (2007) 79.

\bibitem{Dzhunushaliev2008} V.Dzhunushaliev.and V. Folomeev. 6D thick branes from interacting scalar
fields, Phys. Rev. D 77 (2008) 044006. [hep-th/0703043].

\bibitem{wd.goldberger-prl83} W.D.~Goldberger and M.B.~Wise, Modulus stabilization with bulk fields, Phys. Rev. Lett. 83 (1999) 4922.

\bibitem{wd.goldberger-prd60} W.D.~Goldberger and M.B.~Wise, Bulk fields in the Randall-Sundrum compactification
scenario, Phys. Rev. D 60 (1999) 107505.

\bibitem{o.dewolfe-prd62} O.~DeWolfe, D.Z.~Freedman, S.S.~Gubser and A.~Karch, Modeling the fifth-dimension
with scalars and gravity, Phys. Rev. D 62 (2000) 046008.

\bibitem{rn.mohapatra-prd62} R. N.~Mohapatra, A.~Perez-Lorenzana and C. A.~de Sousa Pires, Inflation
in models with large extra dimensions driven by a bulk scalar field, Phys. Rev. D 62 (2000) 105030.

\bibitem{p.kanti-plb481} P.~Kanti, K.A.~Olive and M.~Pospelov, Static solutions for brane models
with a bulk scalar field, Phys. Lett. B 481 (2000) 386.

\bibitem{jm.cline-prd64} J.M.~Cline and H.~Firouzjahi, Brane world cosmology of modulus
stabilization with a bulk scalar field, Phys. Rev. D 64 (2001) 023505.

\bibitem{jm.cline-plb495} J.M.~Cline and H.~Firouzjahi, Five-dimensional warped cosmological solutions
with radius stabilization by a bulk scalar, Phys. Lett. B 495 (2000) 271.

\bibitem{a.flachi-npb610} A.~Flachi and D.~J.~Toms, Quantized bulk scalar fields in the Randall-Sundrum brane model,
Nucl. Phys. B 610 (2001) 144.

\bibitem{m.visser-aip} M. Visser, Lorentzian wormholes from Einsteins to Hawking, AIP Press, (1995).

\bibitem{Gherghetta2000} T. Gherghetta and M. Shaposhnikov. Localizing Gravity on a String-Like Defect
in Six Dimensions. Phys. Rev. Lett. 85 (2000) 240. arXiv:hep-th/0004014v4

\bibitem{Tahim2009} M. O. Tahim, W. T. Cruz, and C. A. S. Almeida. Tensor gauge field localization in branes. Phys. Rev. D 79 (2009) 085022.

\bibitem{Oda2000a} I. Oda. Bosonic fields in the stringlike defect model. Physical Review D 62 (2000) 126009.

\bibitem{Luis-ip} L. S. J. Sousa, W. T. Cruz, and C. A. S. Almeida. Tensor gauge field localization on a string-like
defect, Phys.Lett. B711 (2012) 97.

\bibitem{Liu2007} Y.~X.~Liu, L.~Zhao, X.~H.~Zhang and Y.~S.~Duan, Fermions in self-dual vortex background
on a string-like defect, Nucl. Phys.  B 785 (2007) 234.
\end{thebibliography}
\end{document}